\documentclass[runningheads]{llncs}
\usepackage{graphicx}
\usepackage{amssymb}
\usepackage{amsmath}
\usepackage{hyperref}

\newtheorem{obs}{Observation}
\newtheorem{cor}[theorem]{Corollary}
\newtheorem{result}{Result}

\begin{document}
\title{Reflective Guarding a Gallery}
%
%
\author{Arash Vaezi\thanks{Department of Computer Engineering, Sharif University of 
Technology, {\tt avaezi@ce.sharif.edu}}
        \and
        Bodhayan Roy
        \thanks{Indian Institute of Technology Kharagpur, {\tt broy@maths.iitkgp.ac.in},
    The author is supported by an ISIRD Grant from Sponsored Research and Industrial Consultancy, IIT Kharagpur, and a MATRICS grant from Science and Engineering Research Board.}
        \and
               Mohammad Ghodsi\thanks{Department of Computer Engineering, Sharif University of
Technology, and Institute for Research in Fundamental Sciences
(IPM), {\tt  ghodsi@sharif.edu}} }

\institute{Sharif University of Technology and Indian Institute of Technology Kharagpur}
\maketitle              

\def\OPT{\mbox{\it OPT}}
\def\P{\mbox{\it P}}
\def\NP{\mbox{\it NP}}

\begin{abstract}
This paper studies a variant of the Art Gallery problem in which the ``walls" can be replaced by \emph{reflecting edges}, which allows the guards to see further and thereby see a larger portion of the gallery. 
Given a simple polygon $\P$, first, we consider one guard as a point viewer, and we intend to use reflection to add a certain amount of area to the visibility polygon of the guard. 
We study visibility with specular and diffuse reflections where the specular type of reflection is the mirror-like reflection, and in the diffuse type of reflection, the angle between the incident and reflected ray may assume all possible values between $0$ and $\pi$. 
Lee and Aggarwal already proved that several versions of the general Art Gallery problem are $\NP$-hard.
We show that several cases of adding an area to the visible area of a given point guard are $\NP$-hard, too.

Second\footnote{A primary version of the second result presented here is accepted in EuroCG 2022~\cite{hal-03674221} whose proceeding is not formal.}, we assume that all edges are reflectors, and we intend to decrease the minimum number of guards required to cover the whole gallery.

Chao Xu proved that even considering $r$ specular reflections, one may need $\lfloor \frac{n}{3} \rfloor$ guards to cover the polygon. Let $r$ be the maximum number of reflections of a guard's visibility ray.

In this work, we prove that considering $r$ \emph{diffuse} reflections, the minimum number of \emph{vertex or boundary} guards required to cover a given simple polygon $\cal P$ decreases to
{ $\bf \lceil \frac{\alpha}{1+ \lfloor \frac{r}{8} \rfloor} \rceil$}, where $\alpha$ indicates the minimum number of guards required to cover the polygon without reflection. 
We also generalize the $\mathcal{O}(\log n)$-approximation ratio algorithm of the vertex guarding problem to work in the presence of reflection. 
\end{abstract}
\def\VP{\mbox{\it VP}}
\def\WVP{\mbox{\it WVP}}
\def\LBV{\mbox{\it LBV}}
\def\RBV{\mbox{\it RBV}}
\def\OPT{\mbox{\it OPT}}
\def\InSS{\mbox{\it InSS}}
\def\seg#1{\overline{#1}}
\def\P{\cal P}
\def\T{\cal T}
\def\S{\cal S}
\def\F{\cal F}
\section{Introduction}
Consider a simple polygon $\P$ with $n$ vertices and a point viewer $q$ inside $\P$.  
Suppose $C(\P)$ denotes $\P$'s topological closure (the union of the interior and the boundary of $\P$). Two points $x$ and $y$ are visible to each other, if and only if the line segment $\overline{xy}$ lies completely in $C(\P)$. 
The visibility polygon of $q$, denoted as $\VP(q)$, consists of all points of $\P$ visible to $q$.
Many problems concerning visibility polygons have been studied so far. There are linear-time algorithms to compute $\VP(q)$~(\cite{2}, \cite{1}). Edges of $\VP(q)$ that are not edges of $\P$ are called {\it windows}. 

If some of the edges of $\P$ are made into mirrors, then $\VP(q)$ may enlarge. Klee first introduced visibility in the presence of mirrors in 1969~\cite{3}. 
He asked whether every polygon whose edges are all mirrors is illuminable from every interior point. In 1995 Tokarsky constructed an all-mirror polygon inside which there exists a dark point~\cite{4}.
Visibility with reflecting edges subject to different types of reflections has been studied earlier~\cite{5}: 
(1) \emph{Specular-reflection}: in which the direction light is reflected is defined by the law-of-reflection. Since we are working in the plane, this law states that the angle of incidence and the angle of reflection of the visibility rays with the normal through the polygonal edge are the same.
   (2) \emph{Diffuse-reflection}: that is to reflect light with all possible angles from a given surface. The diffuse case is where the angle between the incident and reflected ray may assume all possible values between $0$ and $\pi$.
 
Some papers have specified the maximum number of allowed reflections via mirrors in between \cite{ad}. In multiple reflections, we restrict the path of a ray
coming from the viewer to turn at polygon boundaries at most $r$
times.
Each time this ray will reflect based on the type of reflection specified in a problem (specular or diffuse).

Every edge of $\P$ can potentially become a reflector. 
However, the viewer may only see some edges of $\P$. When we talk about an edge, and we want to consider it as a reflector, we call it a \emph{reflecting edge} (or a \emph{mirror-edge} considering specular reflections). Each edge has the potential of getting converted into a reflecting edge in a final solution of a visibility extension problem (we use the words ``reflecting edge" and ``reflected" in general, but the word ``mirror" is used only when we deal with specular reflections).

Two points $x$ and $y$ inside $\P$ can see each other through a reflecting edge $e$, if and only if they are reflected visible with a specified type of reflection. We call these points \emph{reflected visible} (or \emph{mirror-visible}). 

\emph{The Art Gallery problem} is to determine the minimum number of guards that are sufficient to see every point in the interior of an art gallery room.
The art gallery can be viewed as a polygon $\P$ of $n$ vertices, and the guards are stationary points in $\P$. If guards are placed at vertices of $\P$, they are called \emph{vertex guards}. 
If guards are placed at any point of $\P$, they are called \emph{point guards}. If guards are allowed to be placed along the boundary of $\P$, they are called \emph{boundary-guards} (on the perimeter).
To know more details on the history of this problem see~\cite{cite:2000}.

The Art Gallery problem was proved to be $\NP$-hard first for polygons with holes by~\cite{cite:rou&sup}. For guarding simple polygons, it was proved to be $\NP$-complete for vertex guards by~\cite{lee}. This proof was generalized to work for point guards by~\cite{cite:aggarval}. 
The class $\exists \mathbb{R}$ consists of problems that can be reduced in polynomial time to the problem of deciding whether a system of polynomial equations with integer coefficients and any number of real variables has a solution. It can be easily seen that $\NP \subseteq \exists \mathbb{R}$. The article~\cite{cite:stoc} proved that the Art Gallery problem is $\exists \mathbb{R}$-complete.
Sometimes irrational coordinates are required to describe an optimal solution \cite{cite:irrational}.

Ghosh~\cite{cite:Ghosh} provided an $\mathcal{O}(\log n)$-approximation algorithm for guarding polygons with or without holes with \emph{vertex} guards. 
King and Kirkpatrick obtained an approximation factor of $\mathcal{O}(\log
\log(\OPT))$ for vertex guarding or perimeter guarding simple polygons~\cite{cite:k}. To see more information on approximating various versions of the Art Gallery problem see \cite{cite:afk}, or \cite{cite:me2020}.

\begin{result}
\label{prob1}
Given a simple polygon $\P$ and a query point $q$ as the position of a single viewer (guard), consider extending the area of the visibility polygon of $q$ ($\VP(q)$) by choosing an appropriate subset of edges and make them reflecting edges so that $q$ can see the whole $\P$.

A) To extend the surface area of $\VP(q)$ by exactly a given amount, the problem is $\NP$-complete.

B) To extend the surface area of $\VP(q)$ using the minimum number of diffuse reflecting edges and by at least a given amount, the problem is $\NP$-hard.
\end{result}

\begin{result}
\label{prob2}
Suppose that in the Art Gallery problem a given polygon, possibly with holes, can be guarded by $\alpha$ vertex guards without reflections, then the gallery can be guarded by at most 
$\lceil \frac{\alpha}{1+ \lfloor \frac{r}{8} \rfloor} \rceil$ guards when $r$ diffuse reflections are permitted. 

For both the diffuse and specular reflection the Art Gallery problem considering $r$ diffuse reflection is solvable in $\mathcal{O}(n^{8^{r+1}+10})$ time with an approximation ratio of $\mathcal{O}( \log n)$.
\end{result}

\subsection{Our Settings}
\begin{figure}[]
\begin{center}
\includegraphics[scale=0.7]{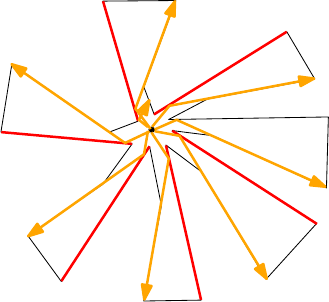}
\caption{This figure illustrates a situation where a single guard is required if we use reflection-edges; $\Theta(n)$ guards are required if we do not consider reflection. Red segments illustrate the reflected-edges.}
\label{fig.example.n}
\end{center}
\end{figure}

Every guard can see a point if the point is directly visible to the guard or if it is reflected visible. This is a natural and non-trivial extension of the classical Art Gallery setting. The problem of visibility via reflection has many applications in wireless networks, and Computer Graphics, in which the signal and the view ray can reflect on walls several times, and it loses its energy after each reflection. 
There is a large literature on geometric optics (such as \cite{optic2}, \cite{optic}, \cite{newton}), and on the chaotic behavior of a reflecting ray of light or a bouncing billiard ball
(see, e.g., \cite{reflection}, \cite{billiard-2}, \cite{billiard-3}, \cite{billiard-4}).
Particularly, regarding the Art Gallery problem, reflection helps in decreasing the number of guards (see Figure~\ref{fig.example.n}).

\setlength{\textfloatsep}{10pt}
\setlength{\intextsep}{10pt}

Sections \ref{exact-reductions} and \ref{sec.atleast} study the problem of extending the surface area of the visibility polygon $\VP(q)$ of a point guard $q$ inside a polygon $\P$ by means of reflecting edges. Section \ref{sec.atleast} considers the scenario in which the visibility polygon of the source needs to be extended at least $k$ units of area where $k$ is a given value. However, to make the problem more straightforward, one may consider adding a specific area with an exact given surface area to the visibility of the source. Section \ref{exact-reductions} considers extending the visibility polygon of $q$ exactly $k$ units of area.

A special reflective case of the general Art Gallery problem is described by Chao Xu in 2011 \cite{ChaoXu}. Since we want to generalize the notion of guarding a simple polygon, if the edges become mirrors instead of walls, the light loses intensity every time it gets reflected on the mirror. Therefore after $r$ reflections, it becomes undetectable to a guard. Chao Xu proved that regarding multiple specular reflections, for any $n$, there exist polygons with $n$ vertices that need $\lfloor \frac{n}{3} \rfloor$ guards. 
Section~\ref{sec:difref} deals with the same problem but regarding diffuse reflection. G. Barequet et al.~\cite{BAREQUET2016123} proved the minimum number of diffuse
reflections sufficient to illuminate the interior of any simple polygon with $n$ walls from any interior-point light source is $\lfloor \frac{n}{2} \rfloor -1$. 
E. Fox-Epstein et al.~\cite{diffuse2} proved that to make 
a simple polygon in a general position visible for a single point light source, we need at most $\lfloor \frac{n-2}{4} \rfloor$ diffuse reflections on the edges of the polygon, and this is the best possible bound. These two papers consider a \emph{single point viewer}; however, in Section~\ref{sec:difref} we considered helping the art gallery problem with diffuse reflection on the edges of the polygon. So, we want to decrease the minimum number of guards required to cover a given simple polygon using $r$ diffuse reflections. We will prove that we can reduce the optimal number with the help of diffuse reflection. Note that we do not assume general positions for the given polygon.
For more information on combining reflection with the art gallery problem see \cite{tcs}, \cite{walcom}, \cite{euro}, \cite{ad}, and \cite{5}. 
\section{Expanding $\VP(q)$ by \emph{exactly} $k$ units of area}
\label{exact-reductions}
We begin this section with the following theorem, and the rest of the section covers the proof of this theorem. 
\begin{theorem}
\label{thm:main}
Given a simple polygon $\P$, a point $q \in \P$, and an integer $k >0$,
the problem of choosing any number of, say  $l$, reflecting edges of $\P$ in order to expand $\VP(q)$ by \emph{exactly} $k$ units of area is $\NP$-complete in the following cases:

1. Specular-reflection where a ray can be reflected only once.
 
2. Diffuse-reflection where a ray can be reflected any number of times.
\end{theorem}

Clearly, it can be verified in polynomial time if a given solution adds
 precisely $k$ units to $\VP(q)$.
Therefore, the problem is in $\NP$.

Consider an instance of the Subset-Sum problem ($\InSS$), which has $val(1) ,val(2)$ $... ,val(m)$ non-negative integer values, and a target number $\T$. Suppose $m \in \Theta(n)$, where $n$ indicates the number of vertices of $\P$. The Subset-Sum Problem involves determining whether a subset from a list of integers can sum to a target value $\T$.  Note that the variant in which all inputs are positive is NP-complete as well~\cite{kleinberg2006algorithm}.

In the following subsections, we will show that the Subset-Sum problem is reducible to this problem in polynomial time. Thus, we deduce that our problem in the cases mentioned above is $\NP$-complete.

\subsection{$\NP$-hardness for specular reflections}
\label{section.specular.exact}
\begin{figure}[htb]
\begin{center}
\includegraphics[scale=0.6]{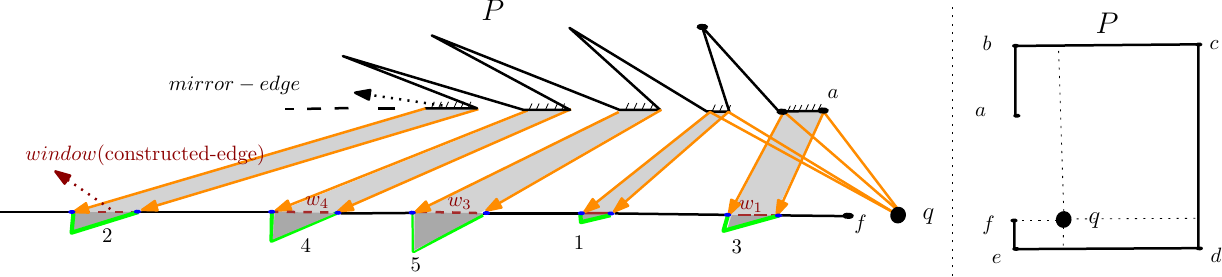}
\caption{Two main components of the reduction polygon is illustrated.} 
\label{fig.1.seperated}
\end{center}
\end{figure}

The reduction polygon $\P$ consists of two rectangular chambers attached side by side. The chamber to the right is taller, while the chamber to the left is shorter but quite broad. The query point $q$ is located in the right chamber (see Figure~\ref{fig.1.seperated}). The left chamber has left-leaning triangles attached to its top and bottom edges. In the reduction from Subset-Sum, the areas of the bottom spikes correspond to the weights of the sets (the values of $\InSS$). The top triangles are narrow and have negligible areas. Their main purpose is 
 to house the edges which may be turned into reflecting edges so that $q$ can see the bottom spikes.

To describe the construction formally, consider $\InSS$.
Denote the $i^{th}$ value by $val(i)$ and the sum of the values till the $i^{th}$ value, $\sum^i_{k=1} val(k)$, by $sum(i)$. We construct the reduction polygon in the following steps:

    (1) Place the query point $q$ at the origin $(0,0)$.
    
    (2) Consider the x-axis as the bottom edge of the left rectangle.
    
  (3)  Denote the left, right and bottom points of the $i^{th}$ bottom spike by $llt(i)$, $rlt(i)$ and $blt(i)$ respectively. Set the coordinates for 
    $llt(i)$ at $(i + 2(sum(i-1)), 0)$,
 $rlt(i)$ at $(i + 2(sum(i)), 0)$, and those for $blt(i)$ at $(i + 2(sum(i)), -1)$.
 
(4) The horizontal polygonal edges between the top triangles are good choices for mirrors, so we call them \emph{mirror-edges}.
        The $i^{th}$ mirror-edge lies between the $(i-1)^{th}$ and $i^{th}$ top spikes. Denote the left and right endpoints of the $i^{th}$ mirror-edge by $lm(i)$ and $rm(i)$ respectively.
        Set the coordinates of $lm(i)$ at $(\frac{(i + 2(sum(i-1)))}{2}, 2(sum(m)+ m))$, and those of $rm(i)$ at $(\frac{(i + 2(sum(i)))}{2}, 2(sum(m)+ m))$.
        
        Denote the topmost point of the $i^{th}$ top spike by $ut(i)$ and set its coordinates at $(\frac{(i + 2(sum(i)))}{2}, 4(sum(m)+ 2m))$.
        

\subsection{Properties of the reduction polygon}
\label{propreties-reduction-polygon}
In this subsection, we discuss properties that follow from the above 
construction of the reduction polygon.
We have the following lemmas.
\begin{lemma}\label{lem:exact}
The query point $q$ can see the region enclosed by the $i^{th}$ bottom spike only through a specular reflection through the $i^{th}$ mirror-edge. 
\end{lemma}

\begin{proof}
See Figure~\ref{fig.rationalco}. The $x$-coordinate of the left and rightmost points of the $i^{th}$ bottom spike is twice that of the $i^{th}$ mirror edge, and the mirror edge is horizontal. Also, the choice of coordinates ensures that the angle $\beta$ is less than the angle $\alpha$. Thus, $q$ cannot see the interior of the $i^{th}$ bottom spike if the $i^{th}$ mirror edge is not chosen as a reflecting edge. Moreover, $q$ sees the whole of the interior of the $i^{th}$ bottom spike if the $i^{th}$ mirror edge is chosen as a reflecting edge.

\setlength{\textfloatsep}{10pt}
\setlength{\intextsep}{10pt}
\begin{figure}[htb]
\begin{center}
\includegraphics[scale=0.8]{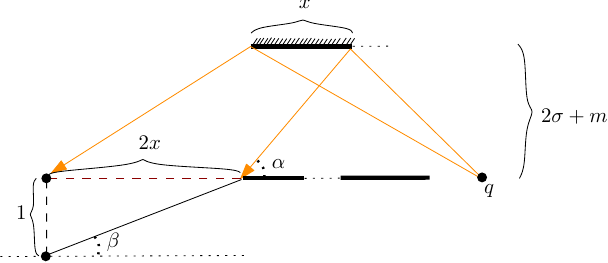}
\caption{This figure illustrates how rational coordinates of the vertices of the spikes are computed and guaranteed.} 
\label{fig.rationalco}
\end{center}
\end{figure}
\end{proof}

\begin{lemma}
\label{lem:rational}
All coordinates of the reduction polygon are rational and take polynomial time to compute.
\end{lemma}
\begin{proof}
This too follows from the construction, as the number of sums used is linear, and each coordinate is derived by at most one division from such a sum.
\end{proof}
\begin{lemma}
The problem of extending the visibility polygon of a query point inside a simple polygon via single specular reflection is $\NP$-complete.
\end{lemma}
\begin{proof}
From Lemma~\ref{lem:exact} it follows that a solution for the problem exists if an only if a solution exists for the corresponding Subset-Sum problem. From Lemma~\ref{lem:rational} it follows that the reduction can be carried out in polynomial time, thus proving the claim.
\end{proof}

\begin{obs}
\label{lem.multiple.specular}
The multiple reflection case of the first case of the problem mentioned in Theorem~\ref{thm:main} is still open. That is the above-mentioned reduction (presented in subsection~\ref{section.specular.exact}) does not work if more than one reflection is allowed.
\end{obs}
\begin{proof}
In the reduction presented in subsection~\ref{section.specular.exact}, each  spike is totally and only visible through one mirror-edge, which is why this reduction and construction does not work in the case of multiple reflections (as an example, the red ray demonstrated in Figure~\ref{fig.1.1} within the polygon reveals that this reduction does not work in the case of multiple reflections). Even if tiny little units of an area from different spikes were partially visible through some other potential mirror-edges (except for their corresponding mirror-edges), their summation could be a large number and the reduction might fail. 

\setlength{\textfloatsep}{10pt}
\setlength{\intextsep}{10pt}
\begin{figure}[htb]
\begin{center}
\includegraphics[scale=0.75]{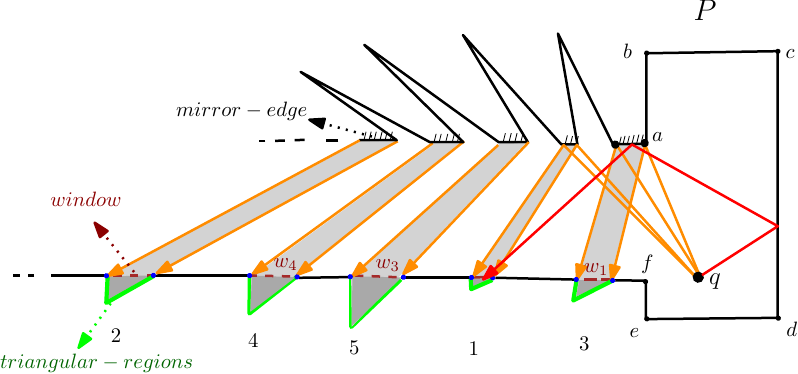}
\caption{The multiple reflection case of the \emph{exact} problem is still open. Every edge has the potential to get converted to a reflector, however, only specified mirror-edges can add as invisible area to the visibility polygon of the viewer. If we consider multiple reflection $\overline{cd}$ can disturb the exclusive functionality of the mirror-edges and the reduction will not work anymore.} 
\label{fig.1.1}
\end{center}
\end{figure}

As Figure~\ref{fig.1.1} shows the multiple reflection rays may disturb the functionality of the mirror-edges. Even after two reflections the segment $\seg{cd}$ may cause a mirror-edge to see some area behind another window. 
On the left side of the polygon the vertical edge of the polygon may disarrange the functionality of some mirror-edge, too.
\end{proof}
\subsection{$\NP$-hardness for diffuse reflections}
\label{dif.reduction}
This subsection deals with the second part of Theorem~\ref{thm:main}. 
Considering diffuse reflection, since rays can be reflected into wrong spikes (a spike which should get reflected visible via another reflecting edge) the previous reduction does not work. Considering multiple plausible reflections, the problem becomes even harder. These rays have to be excluded by an appropriate structure of the polygonal boundary.

The construction presented in this subsection works in the case of multiple reflections, too. Again we reduce the Subset-Sum problem to our problem (Result~\ref{prob1}(A) considering diffuse reflections).

\setlength{\textfloatsep}{10pt}
\setlength{\intextsep}{10pt}
\begin{figure}[htp]
\begin{center}
\includegraphics[scale=0.65]{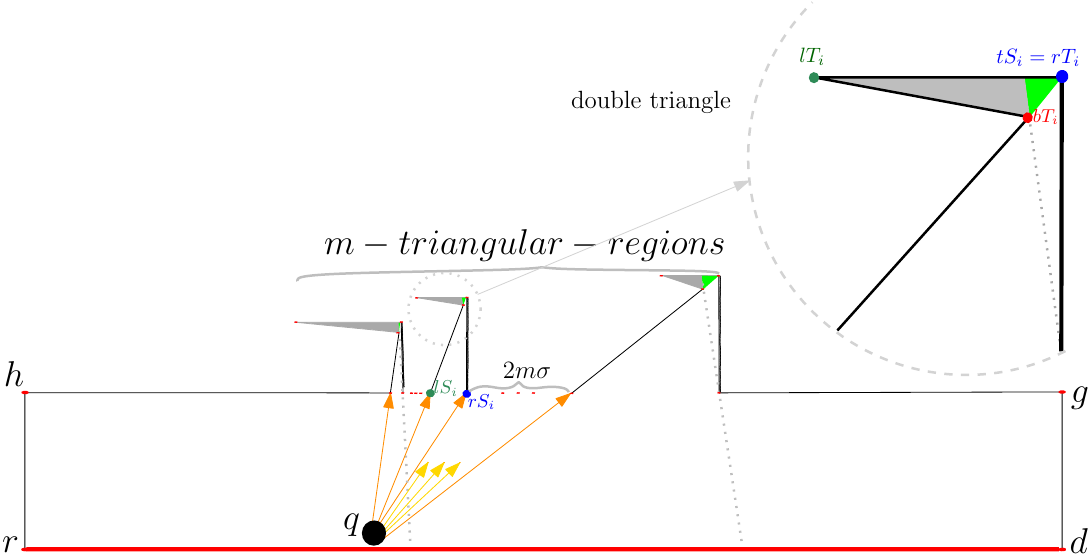}
\caption{A schema of the reduction polygon. Note that $m \in \Theta(n)$. The main polygon is a rectangle, with gadgets on its top edge. 
The green region plus the grey region are in one triangle which its surface area equals to a value of $\InSS$.}
\label{fig:atleast}
\end{center}
\end{figure}

As before, we place the query point $q$ at the origin, $(0,0)$.
The main polygon $\P$ used for the reduction is primarily a big rectangle, with around two-thirds of it being 
to the right of $q$ (see Figure~\ref{fig:atleast}). On top of this rectangle are $m$ \emph{``double triangle''} structures.
Each double triangle structure consists of triangles sharing some of their interiors.
The lower among these two triangles, referred to as the \emph{second triangles}, is right-angled and has its base on
the main rectangle, with its altitude to the right and hypotenuse to the left. The upper triangle,
referred to as the \emph{``top triangle''}, is inverted, i.e., its base or horizontal edge is at the top.
One of its vertices is the top vertex of the second triangle, and another of its vertices merges into the hypotenuse of the second triangle. Its third vertex juts out far to the left, at the same vertical level as the top vertex 
of the second triangle, making the top triangle a very narrow triangle.
The area of the top triangle equals the value of the $i^{th}$ set in the Subset-Sum problem ($val_{i}$).

We make each top triangle to get diffusely reflected visible by only some specific reflecting edges. However, as mentioned previously, there can be a troublemaker shared reflected visible area in each top triangle. This area is illustrated in green in Figure~\ref{fig:atleast}. 
We know that every value of the Subset-Sum problem is an integer. We manage to set the coordination of the polygon so that the sum of the surface area of all the green regions gets equal to a value less than $1$, and all of these regions are entirely reflected visible to the lower edge of the main rectangle. As a result, seeing the green areas through a reflection via the bottom red edge cannot contribute towards seeing exactly an extra region of $k$ units of area. Remember that $k$ is an integer.

Formally, denote the $i$th top and second triangles by $T_{i}$ and $S_{i}$ respectively.
Denote the top, left and right vertices of $S_{i}$ by $tS_{i}$, $lS_{i}$ and $rS_{i}$ respectively.
Denote the sum of values of all subsets of the Subset-Sum problem by $\sigma$. In fact, $\sigma$ = $\sum_{i=1}^{m}val_{i}$.
In general, the triangle $S_{i}$ has a base length of $i$ units, and its base is $2m^{2}\sigma$ units distance away
from those of $S_{i-1}$ and $S_{i+1}$. Therefore the coordinates of $lS_{i}$ and $rS_{i}$ are given by 
$((2m^2\sigma)\frac{i(i+1)}{2} - i, m^2(m+1) \sigma)$ and $(  (2m^2\sigma )\frac{i(i+1)}{2}, m^2(m+1) \sigma )$ respectively.
For any vertex $v$ of the reduction polygon, let us denote the 
$x$ and $y$ coordinates of $v$ by $x(v)$ and $y(v)$ respectively.
The vertex $tS_{i}$ is obtained by drawing a ray originating at $q$ and passing through $lS_{i}$,
and having it intersect with the vertical line passing through $rS_{i}$. 
This point of intersection is $tS_{i}$ with coordinates 
$(x(rS_{i}),m^2(m+1)\sigma+\frac{m^2(m+1)\sigma}{x(lS_{i})})$.

Denote the leftmost and bottom-most vertices of $T_{i}$ by $lT_{i}$ and $bT_{i}$ respectively.
Recall that $bT_{i}$ lies on the hypotenuse of $S_{i}$, and $lT_{i}$ has the same $y$-coordinate as $tS_{i}$.
Moreover, we place $bT_{i}$ in such a way, that the sum of the total regions of all top triangles
seen from the base of the main rectangle (the $rd$ edge in Figure~\ref{fig:atleast}) is less than $1$. 
Intuitively, $bT_{i}$ divides the hypotenuse of $S_{i}$ in the $m^2-1:1$ ratio.
Accordingly, the coordinates of $bT_{i}$ 
are:
$(1+\frac{i(i-1)}{2}+(m^2-1)\frac{(x(tS_{i})-x(lS_{i}))}{(m^2)},m+(m^2-1)\frac{(y(tS_{i})-y(lS_{i}))}{(m^2)})$.

Next, we set coordinates of $lT_{i}$ in a way that the total surface area of $T_{i}$ gets equal to the value of the $i^{th}$ subset. Denote the value of the $i^{th}$ subset by $val_{i}$.
Then, the coordinates of $lT_{i}$ are given by $(x(tS_{i})-2\frac{val_{i}}{y(tS_{i})-y(bT_{i})},y(tS_{i}))$.
Finally, the coordinates of the four vertices of the main rectangle holding all the double triangle gadgets, are given by 
$(-x(rS_{m}),m^2(m+1)\sigma)$, $(-x(rS_{m}),-1)$, $(2(x(rS_{m})),-1)$
and 
$(2(x(rS_{m})), m^2(m+1)\sigma)$.
\begin{lemma}
\label{single.dif.lemma}
The reduction stated in subsection~\ref{dif.reduction} proves that the problem of adding exactly $k$ units of area to a visibility polygon via only a single diffuse-reflection per ray, is $\NP$-complete. The reduction polygon has rational coordinates with size polynomial with respect to $n$.
\end{lemma}
\begin{proof}
 We consider a Subset-Sum problem $\InSS$ on $m$ subsets, and construct a polygon denoted by $\P$ as mentioned in Section~\ref{dif.reduction}. 
 
 Firstly, it is easy to see that the coordinates are rational and
 polynomial in terms of $m$. Note that $m = \Theta(n)$.
 This is because $\InSS$ has a representation linear in $m$, and each double triangle gadget is derived by computing a sum of order $m^4 \InSS$,
 and using only a constant number of the intersection of lines or rays passing through such points. Therefore, polygon $\P$ can be constructed in polynomial time.
 
 Before proceeding further, we prove that the area of $T_{i}$ is $val_{i}$ 
 units of area.
 The length of the base of $T_{i}$ is 
 $\mid x(lt_{i}) - x(tS_{i}) \mid$, and its altitude is 
 $\mid y(tS_{i}) - y(bT_{i})  \mid$.
  So, the area of $T_{i}$ is given by 
  
  $\frac{1}{2}\mid (x(lt_{i})-x(tS_{i}))(y(tS_{i})-y(bT_{i}))\mid$
  
  $=\frac{1}{2}\mid(x(tS_{i})-2\frac{val_{i}}{(y(tS_{i})-y(bT_{i}))}-x(tS_{i}))(y(tS_{i})-y(bT_{i}))\mid$
  
  $=val_{i}$.

 To show that $\P$ is a simple polygon, it is enough to show that no top triangle $T_{i}$
 intersects with $T_{j}$ or $S_{j}$ where $i \neq j$.
 Due to the relative positions of consecutive double triangles, it suffices to show 
 that the $x$ coordinate of $lT_{i}$ is greater than the $x$ coordinate of $tS_{i-1}$.
 To show this, we first prove that the height of each second triangle is greater than $1$.
 Due to its coordinates, triangle $S_{i}$ is similar to the triangle formed by the points $lS_{i}$,
 $q$ and the point with coordinates $(x(lS_{i}), 0)$.
 Therefore the altitude of $S_{i}$ is 
 
 $i(m^{2}(m+1) \sigma) /(  (2m^2\sigma)\frac{i(i+1)}{2} - i)$
 
 $= m/(i + 1 - \frac{2}{m^2 \sigma})$
  which is always at least $1$. 
 
 This means that the altitude of $T_{i}$ is at least $\frac{1}{m^{2}}$, implying that the length of the base of $T_{i}$ is at most $(2m^2)val_{i}$. Since the right vertices 
 of consecutive second triangles are a distance $2m^2\sigma$ units apart,
 this implies that $T_{i}$ can never intersect with $T_{i-1}$ or $S_{i-1}$.
 
 We now show that if there is a solution for a given $k$ in $\InSS$, then
  there is a solution for $k$ in $\P$, i.e. it is possible to extend the visibility
  polygon of $q$ in $\P$ by exactly $k$ units through single reflections of rays.
  This follows readily from the fact (proved above) that each $T_{i}$ has an area of $val_{i}$,
  because due to this, if the $i^{th}$ subset is chosen in a solution of $\InSS$,
  we can make the altitude of $S_{i}$ a reflective edge in the corresponding solution for $\P$,
  and have $q$ see an extra area of $k$ units of area.

 We finally show the opposite direction, i.e. if there is no solution for a given $k$
 in $\P$, then there will be no solution for $k$ in $\InSS$ either.
 To show this, it suffices to prove that the sum over the portions of all top triangles seen by the base of the main rectangle, is a fraction less than one. This works because the only 
 regions that $q$ does not see directly are the top triangles,
 and there are only two ways of $q$ seeing the top triangles;
 either $q$ sees them whole via a reflection on the altitude of the corresponding second triangle,
 or just partially via a reflection on the base of $\P$.
  The left and right sides of the main rectangle constituting $\P$ cannot see any $T_{i}$
 due to their faraway $x$-coordinates.
 So, 
 excluding the unique and trivial case where the altitude of each $S_{i}$ is chosen,
 choosing the base of $\P$
 as a reflecting edge always yields a fractional value of the extra area added, which contradicts
 the integral value of $k$.
 
 Consider any $S_{i}$. Let the point on the altitude of $S_{i}$ sharing the same y coordinate as 
 $bT_{i}$ denoted by $a_{i}$. Consider the small triangle formed by these two points and $tS_{i}$
 and denote it by $U_{i}$. Naturally, the base of $U_{i}$ is parallel to the base of $S_{i}$,
 since both are parallel to the x - axis. Again, since the ratio between $\overline {tS_{i} bT_{i}}$
 and $\overline {tS_{i} ls_{i}}$ is $1: m^{2}$, the ratio between the areas of $U_{i}$ and $S_{i}$ 
 must be $1: m^{4}$. 
 
 Our computations above show that the area of any $S_{i}$ is $\frac{mi}{2(i + 1 - \frac{2}{m^2 \sigma})}$,
 upper bounded by $n/2$.
 Therefore the total
 area of $U_{i}$ for all second triangles is upper bounded by $m (\frac{1}{m^4}) (\frac{m}{2}) = \frac {1}{2m^2}$. It follows from basic geometrical considerations that for each $S_{i}$,
 the region of $T_{i}$ seen by the base of $P$ is upper bounded by the region of $T_{i}$
 cut off by the ray originating at $rS_{i}$ and passing through $bT_{i}$, and hence, 
 twice the area of $U_{i}$. So, the sum total of the regions of the top triangles
 seen by the base of $\P$ cannot exceed $\frac{1}{m^{2}}$. This concludes the reduction.
\end{proof}

We can use the above-mentioned reduction in case multiple reflection is allowed. 
See the following corollary.

\begin{cor}
\label{cor.redu2.num}
The reduction stated in subsection~\ref{dif.reduction} works if multiple reflections is allowed.
\end{cor}
\begin{proof}
Use the reduction stated in the proof of Lemma~\ref{single.dif.lemma}.
Call the reflecting edges ($\overline{tS_{i}rS_{i}}$), as the main-reflecting edges. To expand $\VP(q)$ by exactly $k$ units of area, any solution needs to contain a subset of these main-reflecting edges, and this subset exists if and only if there exists an equivalent subset in $\InSS$. No other edge can make an invisible area reflected visible to the viewer except for main-reflecting edges. 


\paragraph{Multiple Reflections}
See Figure~\ref{fig.x.point} consider the edge ($\overline{af}$) that has the point $x$ on it. In fact, the edge is equivalent to $\overline{tS_{i}rS_{i}}$. We denote this edge as the \emph{``second-reflecting edge"}. Every green area includes a part of this edge ($\seg{ax}$).  The segment $\seg{ax}$ is not directly visible to $q$. A triangular spike is entirely reflected visible to both its corresponding main-reflecting edge and the second-reflecting edge.

\begin{figure}[htp]
\begin{center}
\includegraphics[scale=0.7]{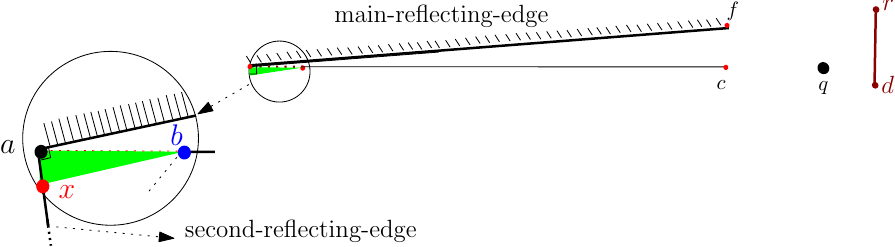}
\caption{This figure illustrates one double triangle structure. The double triangle structures are placed on the top edge of the main rectangle illustrated in Figure~\ref{fig:atleast}. In the multiple reflection case of the problem we should count on two edges as the reflecting edges: main-reflecting edge and second-reflecting edge.
The green part is the area that is visible through the $\overline{rd}$ edge. }
\label{fig.x.point}
\end{center}
\end{figure}

Consider two reflections, $q$ can see a triangular spike through the $\seg{rd}$-edge (the bottom edge of the main rectangle in Figure~\ref{fig:atleast}) and its corresponding second-reflecting edge in a double triangular gadget. Therefore, considering more than one reflections, $q$ can see a triangular spike either by a main-reflecting edge, or a second-reflecting edge. 

Now, we can use a reduction similar to the one used in the single reflection case (in the proof of Lemma~\ref{single.dif.lemma}), except that here we need to count on both the main-reflecting edge and the second-reflecting edge. Considering a solution, if either one of these reflecting edges is selected, then we need to choose the corresponding value of $\InSS$.


Note that to make a corresponding selection of edges to cover a specific spike and to assign those edges to the values of the subset sum is what we had in the reductions. Therefore, no matter how many reflections are allowed, we have to make a specific set of edges into reflectors to cover a specific spike. Also, keep in mind that when rays cover a whole edge, no matter how many sources are covering that edge, the diffuse reflection of that edge does not change. The angles of the sources are not important, unlike the specular reflection. That is to say, if an edge is covered by rays from a source, adding more rays from other sources does not add anything to the diffuse reflection of that edge. This feature of diffuse reflection makes the reductions work in multiple reflections, too. So, an appropriate subset of edges corresponds to a solution of the instance of the subset sum problem. 
\end{proof}
\section{Expanding at least $k$ units of area}
\label{sec.atleast}
In this section, we modify the reduction of Lee and Lin~\cite{lee} and use it to infer that the problem of extending the visibility polygon of a given point by a region of area at least $k$ units of area with the \emph{minimum} number of reflecting edges is $\NP$-hard, where $k$ is a given amount (Result~\ref{prob1}(B)). 
The idea is that the potential vertex guards are replaced with edges that can reflect the viewer ($q$). We need an extra reflecting edge, though.
Only a specific number of edges can make an invisible region (a spike) entirely reflected visible to the viewer. Converting the correct minimum subset of these edges to the reflecting edges determines the optimal solution for the problem.

The specular-reflection cases of the problem are still open. Nonetheless, it was shown by Aronov~\cite{5} in 1998 that in such a polygon where all of its edges are mirrors, the visibility polygon of a point can contain holes. And also, when we consider at most $r$ specular reflections for every ray, we can compute the visibility polygon of a point inside that within $\mathcal{O}(n^{2r} \log n)$ of time complexity and $\mathcal{O}(n^{2r})$ of space complexity~\cite{ad}.

\begin{conjecture}
\label{t.7}
Given a simple polygon $\P$, and a query point $q$ inside the polygon, and a positive value $k$, the problem whether $l$ of the edges of the polygon can be turned to reflecting edges so that the area added to $\VP(q)$ through (single/multiple) diffuse-reflections increases at least $k$ units of area is $\NP$-hard. 
\end{conjecture}
\emph{Proof Idea.}
First, see the following definition:

\emph{The 3-SAT problem} is defined like this; Given a boolean formula in conjunctive normal form, with each clause having exactly three variables, can any assignment of the variables satisfy the formula? We assume that clauses with one variable and its complement, and also clauses with two variables have already been removed. Here, we consider an instance of this problem with $\xi$ variables and $\gamma$ clauses.

The construction uses the method of reduction of the 3-SAT problem to the Art gallery problem with vertex guards, by Lee et al~\cite{lee}. In the original reduction, for a given 3-SAT formula with $\xi$ variables and $\gamma$ clauses, a simple polygon was constructed, which would require a minimum of exactly $\xi + 3\gamma + 1$ vertex guards to be completely seen if and only if the given 3-SAT formula was satisfiable.

\begin{figure}[htb]
\begin{center}
\includegraphics[scale=0.7]{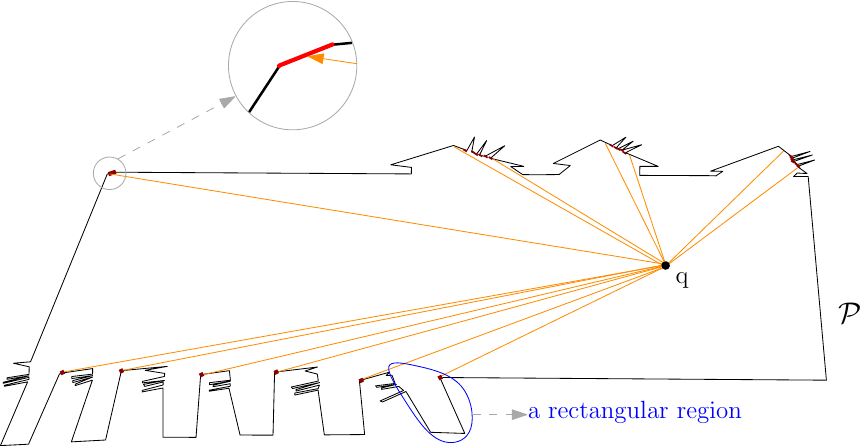}
\caption{An example for the Modified-Art-Gallery-polygon. The red edges are reflecting edges. These reflecting edges can reflect light only to their front side.}
\label{fig:modified}
\end{center}
\end{figure}

In the original construction, there are several spikes, for which it would be
necessary to turn certain vertices into vertex guards if the whole polygon is to be
made visible from only $\xi+3\gamma+1$ vertices. We call these the \emph{candidate-vertices}.
In our construction, we replace the candidate vertices for vertex guards with small edges that we call \emph{reflecting edges}.
Note that the same as in the original construction, certain vertices are necessary to be turned into vertex guards; here, it is necessary to turn certain edges into reflecting edges.
We can do this by adding a suitably small $\epsilon$ to the coordinates of every candidate vertex and moving on the edges of the polygon to add an edge with at most $2\epsilon$ length, such that both its end-vertices are seen by the query point (the viewer).
These are marked in red in Figure~\ref{fig:modified}.
We also need an extra reflecting edge at the top-left corner of the polygon, required to see the rectangular regions that contain the variable gadgets on the bottom of the polygon so that the query point $q$ sees the whole surfaces of all the rectangular regions. Note that every diffuse reflecting edge can only see a half-plane on its front side. So, the region behind this half-plane in every rectangular region must be covered by other reflectors. 
To explain more, note that the position of $q$ and the vertex guards specified in the original reduction have already made the clause gadgets on top and the spikes in the variable gadgets in the bottom directly or reflected visible to $q$. However, because of the angular direction and the position of $q$, the main rectangular regions in the variable gadgets cannot be directly or reflected visible to $q$, so we need an extra reflector. All the rectangular regions are visible to the left top vertex of the central rectangle, so we replaced that vertex with a diffuse reflector. Here, we do not give details of the original decomposition. Instead, we refer interested readers to \cite{lee} for further details.  
We set $k$ to be the difference between the area of the polygon and the area of $\VP(q)$. Based on the analysis used in~\cite{lee}, this makes the polygon to require $l = \xi + 3\gamma+1$ reflecting edges to be covered if and only if the given 3-SAT formula is satisfiable.



\section{Regular Visibility vs Reflection}
\label{sec:difref}

\begin{figure}[htb]
\begin{center}
\includegraphics[scale=0.4]{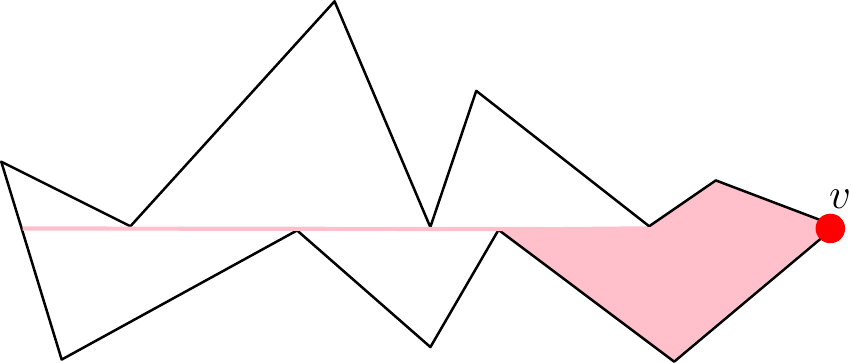}
\caption{The visibility polygon of a point $v$.} 
\label{fig:vp}
\end{center}
\end{figure}

This section deals with Result~\ref{prob2}.
Under some settings, visibility with reflections can be seen as a general case of regular visibility. For example, consider guarding a polygon $\P$ with vertex guards, where all the edges of the polygons are diffuse reflecting edges, and $r$ reflections are allowed for each ray.
Let $S$ be the set of guards in an optimal solution if we do not consider reflection.
Since we allow multiple vertices of the polygon to be collinear, we slightly change the notion of a visibility polygon for convenience. Consider the visibility polygon of a vertex (see Figure~\ref{fig:vp}). It may include lines containing points that do not have any interior point of the visibility polygon within a given radius. So, given a vertex $x$ of $\P$, we consider the union of the interior of the original $\VP(x)$ (denoted by $int\VP(x)$), and the limit points of $int\VP(x)$, as our new kind of visibility polygon of $x$. Clearly, a guard set of $\P$ gives a set of the new kind of visibility polygons whose union is $\P$.

\setlength{\arrayrulewidth}{0.5mm}
\setlength{\tabcolsep}{10pt}

Consider any guard $v \in S$. The visibility polygon $\VP(v)$ of $v$ must have at least one window\footnote{A window is an edge of a viewer's visibility polygon, which is not a part of an edge of the main polygon.}. Otherwise, $v$ is the only guard of $\P$. Consider such a window, say, $w$. 
Let $x$ be a point of intersection of $w$ with the polygonal boundary.
Then there must be at least another guard $u \in S$ such that $x$ lies in both $\VP(u)$ and $\VP(v)$. 
The following lemmas discuss how $\VP(u)$ and $\VP(v)$ 
can be united using a few diffuse reflections, and how the whole polygon can be seen by a just a fraction of the optimal guards, depending on the number of reflections allowed per ray.

\begin{theorem}
\label{thm:bnd}
If $\P$ can be guarded by $\alpha$ vertex guards without reflections, then $\P$ can be guarded by at most 
$\lceil \frac{\alpha}{1+ \lfloor \frac{r}{8} \rfloor} \rceil$ guards when $r$ diffuse reflections are permitted.
\end{theorem}
\label{proof.thm:bnd}
To prove this theorem see the following lemmas first:

\begin{lemma} 
\label{lem:2ref}

If $\mathcal{S}$ is an optimal vertex guard set of polygon $\P$ and $|\mathcal{S}|>1$
then for every guard $u \in \mathcal{S}$ there exists a different guard $v \in
\mathcal{S}$" such that $u$ and $v$ can see each other through five diffuse reflections. Furthermore, 
$u$ and $v$ can fully see each other's visibility polygons with eight diffuse reflections.
\end{lemma}
\begin{proof}
If $u$ sees $v$, then the proof follows easily from the general case where $u$ does not see any guard $v$, so we assume the latter.  
Since $u$ does not see the whole of $\P$, there must exist a guard $v$ and a polygonal boundary point $x$, such that the boundaries of $\VP(u)$ and $\VP(v)$ intersect at $x$.
If $x$ is not a polygonal vertex, then $u$ sees $v$ via one reflection. 
If $x$ is a polygonal vertex, and $u$ and $v$ both see a point $y$, which is not a polygonal vertex, then again, only one diffuse reflection is required.
Otherwise, let $a$ and $b$ be two arbitrarily close to $x$ and on the polygonal boundary on both sides of $x$. 
Consider any point $c$ seen by $x$, such that $c$ is not a polygonal vertex and lies in the interior of the wedge formed by the rays $\overrightarrow{bx}$ and $\overrightarrow{ax}$. 


Suppose that no other vertex of $\VP(u)$ lies on $\overline{ua}$,
no other vertex of $\VP(v)$ lies on $\overline{vb}$.
Then, $a$ and $b$ can be used for reflection by $u$ and $v$, respectively.
So, $u$ sees $v$ via three reflections through, $a$, $c$ and $b$ in the same order.

Now consider the other case where
a vertex of $\VP(u)$ lies on $\overline{ua}$,
or a vertex of $\VP(v)$ lies on $\overline{vb}$. Without loss of generality, let a vertex of $\VP(u)$ lies on $\overline{ua}$.
Then consider the ray $\overrightarrow{ua}$. Using $u$ as a pivot, rotate $\overline{ua}$ away from the polygonal boundary by an arbitrarily small angle $\theta$ to hit a non-vertex polygonal boundary point (denoted by $z$).
Clearly, $z$ is visible from both $u$ and $a$. So we can reach $a$ from $u$ via a single reflection through $z$. A similar reasoning between $v$ and $b$ gives another extra reflection, say, through a point $y$.
Thus, $u$ can always see $v$ via at most five reflections.

Now, consider the vertex $v$ and its visibility polygon (any point $q$ seen by $v$). On the two polygonal boundary edges incident to $v$, consider two points $g$ and $h$ arbitrarily close to $v$. As before (as in case for $x$), we can find a non-vertex point $j$ on the polygonal boundary, so that $g$ can see $h$ via a reflection through $j$. Since $g$ and $h$ are arbitrarily close to $v$, any point in $VP(v)$ is visible from $g$ or $h$. Since $v$ is visible from $y$, either $g$ or $h$, or both, are visible from $y$. Then $g$, $j$ and $h$ are three more points through which we can reflect to see the whole of $\VP(v)$ from $u$, giving  total of at most eight reflections (see Figure~\ref{fig:vpr}). 

\setlength{\textfloatsep}{10pt}
\setlength{\intextsep}{10pt}
\begin{figure}[htb]
\begin{center}
\includegraphics[scale=0.5]{rvis3.pdf}
\caption{In this example, $\VP(v)$ is visible from $u$ via at most eight diffuse reflections. In this figure, $q\in \VP(v)$, and $u$ sees $q$ via reflections through $z$, $a$, $c$, $b$, $y$, $g$, $j$, $h$, in that order.
} 
\label{fig:vpr}
\end{center}
\end{figure}

\end{proof}

Now we build a graph $G$ as follows. We consider the vertex guards in $\mathcal{S}$ as the vertices of $G$, and add an edge between two vertices of $G$ if and only if the two corresponding vertex guards in $\mathcal{S}$ can see each other directly or through at most five reflections. We have the following Lemma.
\begin{lemma} 
\label{lem:con}
The graph $G$ is connected.
\end{lemma}
\begin{proof}
Consider any guard $g_{i}$ of $\mathcal{S}$. Let $g_{j}$ be the clockwise next guard on the polygonal boundary. Suppose $\VP(g_i)$ and $\VP(g_j)$ have a common point of intersection on the polygonal boundary. Then clearly, there is a path in $G$ between $g_i$ and $g_j$.
Otherwise, suppose there is no such point of intersection. Traverse clockwise on the polygonal boundary till $\VP(g_i)$ ends. The point on the polygonal boundary where  
$\VP(g_i)$ ends, must be seen by some other guard $g_k$, which is adjacent to $g_i$ in $G$. Likewise, we traverse the polygonal boundary till $\VP(g_j)$ is reached. Each guard that occurs so far is adjacent to the previous one in $G$. Therefore $G$ contains a path between $g_i$ and $g_j$. Continuing our traversal throughout the complete polygonal boundary,
we get paths between every pair of vertices of $G$, implying that $G$ is connected.



\end{proof}


Consider any optimal vertex guard set  $\mathcal{S}$ for the Art Gallery problem on the polygon $\P$, where $\mid \mathcal{S} \mid = \alpha$. 
 Build a graph $G$ on $\mathcal{S}$ as it was mentioned before Lemma~\ref{lem:con}.
Due to Lemma \ref{lem:con}, $G$ is connected. Find a spanning tree $T$ of $G$ and root it at any vertex. Denote the $i^{th}$ level of vertices of $T$ by $L_i$. Given a value of $r$, divide the levels of $T$ into $1+ \lfloor \frac{r}{8}\rfloor$ classes,
such that the class $\mathcal{C}_{i}$ contains all the vertices of all levels of $T$ of the form $L_{i+x(1+ \lfloor \frac{r}{8}\rfloor)}$, where $x\in \mathbb{Z}^+_0$.
By the pigeonhole principle, one of these classes will have at most  $\lceil \frac{\alpha}{1+ \lfloor \frac{r}{8} \rfloor} \rceil$ vertices. Again, by Lemma \ref{lem:2ref}, given any vertex class $\mathcal{C}$ of $T$, all of $\P$ can be seen by the vertices of $\mathcal{C}$ when $r$ diffuse reflections are allowed. The theorem follows.
\begin{cor}
The above bound (mentioned in Theorem~\ref{thm:bnd}) holds even if the guards are allowed to be placed anywhere on the boundary of the polygon.
\end{cor}
\begin{proof}
 The proof follows directly from the proof of Theorem \ref{thm:bnd} since Lemmas \ref{lem:2ref} and \ref{lem:con} are valid for boundary guards as well.
\end{proof}
\begin{obs}
\label{obs:nopointguard}
The above bound (mentioned in Theorem~\ref{thm:bnd}) does not hold in the case of arbitrary point guards.
\end{obs}
\begin{proof}
See Figure \ref{fig.pg}. The two guards, colored red, see the whole of the polygons. Clearly, if we seek to replace them with one guard and allow reflections, then the new guard must lie somewhere near the polygon's lowest vertex. From there, it can see both the red guards. On the left and right sides of the red guards are two funnels whose apices are visible only from certain edges of their respective opposite funnels. These edges are not visible to any point near the lowest vertex of the polygon. So, it is impossible for only one guard to see the whole polygon through a constant number of reflections. For any given $k$, the funnels can be made narrower so that   $k$ reflections are not enough to see the whole polygon.

\setlength{\textfloatsep}{10pt}
\setlength{\intextsep}{10pt}
\begin{figure}[htb]
\begin{center}
\includegraphics[scale=0.2]{pointguard.pdf}
\caption{Two point guards cannot be replaced by one despite allowing reflections.} 
\label{fig.pg}
\end{center}
\end{figure}

\end{proof}

Finding an approximate solution to the vertex guard problem with $r$ diffuse reflections is harder than approximating the standard problem. Reflection may change the position of guards remarkably. Here, we have a straight-forward generalization of Ghosh's discretization algorithm presented in~\cite{artgal}.


\begin{theorem}
\label{thm:approx}
For vertex guards, the art gallery problem considering $r$ reflections, for both the diffuse and specular reflection are solvable in $\mathcal{O}(n^{8^{r+1}+10})$ time giving an approximation ratio of $\mathcal{O}( \log n)$.
\end{theorem}
\begin{proof}

We begin by drawing all possible windows and decomposing the polygon into convex polygons, as it is mentioned in \cite{artgal}. 
Denote the set of these convex polygons by $R_0$. Denote the set of vertices of these convex polygons of $R_0$, that lie on the boundary of $\P$ by $Q_0$. Note that in the initial step, zero diffuse reflections are considered. In the next step, allow five diffuse reflections for each ray. Each vertex's visibility polygons are extended and have new points on the boundary of $\P$ as their vertices. Join all such pairs of vertices, whenever possible, by drawing windows, to get a new larger set of convex polygons. Denote the set of convex polygons so obtained by $R_1$, and the set of all their vertices lying on the boundary of $\P$ by $Q_1$. 
Analogously, we associate with $i$ diffuse reflections the sets $R_i$ and $Q_i$.

For $r$ diffuse reflections, as before, draw all possible windows from $Q_{r-1}$ and compute the minimal convex polygons formed as a result. 
Start by updating $R_{r-1}$ to $R_r$ and then update $Q_{r-1}$ to $Q_r$ by updating their constituent convex polygons and their vertices lying on the boundary of $\P$, respectively. The lines in $R_r$ are formed by joining together the end-points of the lines of $R_{r-1}$. So, the cardinality of $R_r$
is at most $8\lvert R_{r-1} \rvert^{10}$. The cardinality
of $Q_r$ is at most $\lvert R_{r} \rvert^{10}$, due to being formed 
by the intersections of the lines of $R_r$. Note that the cost of computing each reflection is two, and we are computing five successive reflections, where each takes quadratic time.

Hence, by the argument in Theorem~2.1 of \cite{artgal}, the algorithm takes a total time of $\mathcal{O}(n^{8^{r+1}+10})$. The extra exponent of $10$ comes due to traversal following the method of \cite{artgal} again.
Our algorithm also gives an approximation ratio of $\mathcal{O}(\log n)$ 
due to being reduced from the greedy algorithm of Set Cover \cite{cite:Ghosh,artgal}.

\end{proof} 

\section{Conclusion}
In this paper, we deal with a variant of the Art Gallery problem in which the guards are empowered with reflecting edges. 
Many applications consider one source and they want to make that source visible via various access points. Consider a WiFi network in an organization where due to some policies all the personnel should be connected to one specific network. The access points should receive the signal from one source and deliver it to places where the source cannot access. The problem is to minimizing the access points.

The gallery is denoted by a given simple polygon $\P$. 
This article mentioned a few versions of the problem of adding an area to the visibility polygon of a given point guard inside $\P$ as a viewer. Although we know that reflection could be helpful, we proved that several versions of the problem are $\NP$-hard or $\NP$-complete. 

Nonetheless, we proved that although specular reflection might not help decrease the \emph{minimum} number of guards required for guarding a gallery, diffuse reflection can decrease the optimal number of guards.



\bibliographystyle{ieeetr}
\bibliography{mybibliography.bib}

\end{document}